\documentclass{iau}
\usepackage{graphicx,hyperref,natbib}

\title[Evolution of binary supermassive black holes]
{Evolution of binary supermassive black holes \\ and the final-parsec problem}

\author{Eugene Vasiliev}
\affiliation{Lebedev Physical Institute, Moscow, Russia\\ 
email: {\tt eugvas@lpi.ru}}

\pubyear{2015}
\volume{312}
\pagerange{1--1}
\jname{Star clusters and black holes in galaxies across cosmic time}
\editors{R.~Spurzem \& F.-K.~Liu, eds.}
\begin{document}

\maketitle

\begin{abstract}
I review the evolution of binary supermassive black holes and focus on the stellar-dynamical 
mechanisms that may help to overcome the final-parsec problem -- the possible stalling of 
the binary at a separation much larger than is required for an efficient gravitational wave 
emission. Recent $N$-body simulations have suggested that a departure from spherical symmetry 
in the nucleus of the galaxy may keep the rate of interaction of stars with the binary 
at a high enough level so that the binary continues to shrink rather rapidly. 
However, a major problem of all these simulations is that they do not probe the regime where 
collisionless effects are dominant -- in other words, the number of particles in 
the simulation is still not sufficient to reach the asymptotic behaviour of the system. 
I present a novel Monte Carlo method for simulating both collisional and collisionless 
evolution of non-spherical stellar systems, and apply it for the problem of binary 
supermassive black hole evolution. I show that in triaxial galaxies the final-parsec problem 
is largely non-existent, while in the axisymmetric case it seems to still exist in the limit 
of purely collisionless regime relevant for real galaxies, but disappears in the $N$-body 
simulations where the feasible values of $N$ are still too low to get rid of collisional 
effects.
\end{abstract}

\firstsection
\section{Introduction}

Most galaxies are believed to host central massive black holes (MBH) \citep{FerrareseFord2005,
KormendyHo2013}.
In the hierarchical merger paradigm, galaxies in the Universe typically experience several 
major and multiple minor mergers in their lifetime, and every such merger results in the 
eventual formation of MBH binary. The evolutionary stages of MBH binary were outlined 
in \citet{BegelmanBR1980}: 
\begin{itemize}
\item The merger of two galaxies creates a common nucleus; both MBH sink to the center 
of the merger remnant due to dynamical friction and form a binary (distance $a\sim 10$~pc, 
all estimates are given for a $10^8\,M_\odot$ black hole). The two black holes are 
considered a binary when they start to ``feel'' each other (their mutual gravitational force 
is larger than the force from the rest of the galaxy).
\item Three-body interactions between the MBH binary and the stars of galactic nucleus 
eject stars from the vicinity of the binary, and form 
a depleted zone, or ``mass deficit'' \citep{MilosMRB2002}, in the nucleus 
\citep[which indeed has been identified in several galaxies, see e.g.][]{DulloGraham2014}. 
The initial stage of this process is very rapid and leads to a decrease of the binary's 
semimajor axis $a$ by a factor of few to ten ($a\sim 1$~pc).
\item Once the binary becomes hard and expels almost all stars that ever come close to 
the binary, its subsequent shrinking could slow down considerably, and could even last 
longer than the Hubble time -- a situation that has been called the ``final-parsec problem'' 
\citep{MilosMerritt2003a}.
\item If, despite all troubles, the binary shrinks to a separation $\lesssim 10^{-2}$~pc, 
the emission of gravitational waves (GW) becames the dominant mechanism that extracts 
the energy from the binary, leading to its rapid coalescence ($a \sim 10^{-5}$~pc).
\end{itemize}

A star passing at a distance $\lesssim 2a$ from the binary will experience a complex 
3-body interaction which results in ejection of the star with a larger velocity than 
it had before the interaction (the slingshot effect).
These stars carry away energy and angular momentum from the binary, so that its 
semimajor axis $a$ decreases: $\frac{d}{dt}(1/a) \approx 16 G\,\rho / \sigma
\;\equiv\; H_\mathrm{full}$ \citep{Quinlan1996}.
Thus, if density of field stars $\rho$ remains constant (their velocity dispersion 
$\sigma$ usually does), the binary hardens with a constant rate.
However, the reservoir of low angular momentum stars (so-called loss cone -- stars that 
have small enough pericenter distances and can be ejected) is finite and is depleted 
rather quickly. The subsequent evolution depends on the rate of the loss-cone repopulation.

\section{Loss-cone theory in spherical and non-spherical systems}

The classical loss-cone theory was developed in application to the problem of star capture 
by a single MBH (e.g.\ 
\citealt{FrankRees1976,LightmanShapiro1977}, see also a recent review in \citealt{Merritt2013}).
In the case of a binary MBH, the loss-cone boundary is defined by the radius of binary's 
orbit rather than the radius of capture of tidal disruption (and therefore is much larger).
In terms of angular momentum $L$ of a star, the boundary corresponds to 
$L_\mathrm{LC} \equiv \sqrt{2\kappa\,G\,M_\mathrm{bin}\,a}$, with $\kappa\simeq 1$; 
most stars entering the loss cone are on highly eccentric orbits 
($L_\mathrm{LC} \ll L_\mathrm{circ}$, where the latter quantity is the angular momentum 
of a circular orbit with the same energy).
The stars inside the loss cone are eliminated within one orbital period $T_\mathrm{orb}$. 
The crucial parameter for the evolution is the timescale for repopulation of the loss cone.
In the absence of other processes, the repopulation time is a small fraction of the two-body 
relaxation time:
\begin{equation} \nonumber
T_\mathrm{rep} \sim T_\mathrm{rel}\,\frac{L_\mathrm{LC}^2}{L_\mathrm{circ}^2}
\mbox{, where }\;
T_\mathrm{rel} = \frac{0.34\,\sigma^3} {G^2\, m_\star\, \rho_\star\,\ln\Lambda} .
\end{equation}
If $T_\mathrm{rep} \lesssim T_\mathrm{orb}$, the loss cone is full (refilled faster 
than orbital period). In real galaxies, however, the opposite regime applies -- 
the empty loss cone. In this case the hardening rate is much lower:
$H\equiv \frac{d}{dt}(a^{-1}) \simeq H_\mathrm{full}\,T_\mathrm{orb}/{T_\mathrm{rep}}$.
Relaxation is too slow for an efficient repopulation of the loss cone: 
in the absense of other processes the binary would not merge in a Hubble time.

There are several possible ways to enhance the loss-cone repopulation and hardening rate 
of the binary:
\begin{itemize}
\item \textit{Brownian motion} of the binary enables its interaction with a larger number 
of stars \citep{MilosMerritt2001, ChatterjeeHL2003, MilosMerritt2003b}.
\item \textit{Time-dependent solution} of the Fokker-Planck equation describing 
the diffusion of stars in angular momentum generally yields a higher loss-cone 
repopulation rate at the initial stage of evolution due to steeper gradients in 
the phase space distribution of stars \citep{MilosMerritt2003b}.
\item \textit{Secondary slingshot} allows stars scattered by the binary with velocities 
lower than escape velocity to return to the galaxy center and interact with the binary 
more than once \citep{MilosMerritt2003b}, although this effect seems to be suppressed 
in non-spherical systems due to non-radial trajectories of ejected stars 
\citep{VasilievAM2014}.
\item \textit{Gas} in the circumbinary accretion disk may increase the dissipation of 
binary's binding energy, however, its effect may be rather difficult to model and it 
is not guaranteed to solve the hardening problem \citep{LodatoNKP2009, Roskar2014}; 
moreover, this factor does not apply for dry mergers (gas-poor galaxies) which are 
the subject of this study.
\item \textit{Perturbations} from a possible third black hole \citep[e.g.][]{IwasawaFM2006, 
AmaroSeoane2010} may increase the binary eccentricity due to the Kozai-Lidov mechanism, 
thereby driving it into GW-dominated regime sooner; 
other massive objects such as giant molecular clouds \citep{PeretsAlexander2008} may 
substantially increase the two-body relaxation rate and efficiently scatter stars into 
the loss cone.
\item \textit{Non-spherical} shape of the merger remnant gives rise to torques that 
change stars' angular momenta and have been proposed as a possible mechanism that keeps 
the loss cone full \citep{Yu2002, MerrittPoon2004, HolleySigurdsson2006}. 
\end{itemize}

Let us explain the last point in more detail. 
If the distribution of stars and the gravitational potential that they generate 
is not exactly spherical, then the angular momentum $L$ of any star is not conserved, 
but experiences oscillations due to torques from the non-spherical potential. 
The period of these oscillations is a few times $T_\mathrm{orb}$, and their amplitude 
depends on the degree of flattening, as well as on the average value of $L$ (stars with 
$L\ll L_\mathrm{circ}$ tend to have larger variations in $L$). 
Therefore, due to non-spherical torques, a much larger number of stars can attain low 
values of $L$ and enter the loss cone at some point in their (collisionless) evolution, 
regardless of two-body relaxation. 
In the axisymmetric potential, one component of $L$ parallel to the symmetry axis ($L_z$) 
is conserved, therefore the total angular momentum cannot become smaller than $L_z$; 
still, there are many more stars with $L_z<L_\mathrm{LC}$ than stars in the loss cone 
(with the instantaneous value of $L<L_\mathrm{LC}$), and a large fraction of them can 
eventually attain $L<L_\mathrm{LC}$. In the triaxial case, there is an entire population 
of centrophilic orbits such as pyramids and chaotic orbits 
\citep[e.g.][Figure~1g,a]{PoonMerritt2001}, 
that can achieve arbitrary low values of $L$ -- all of them could 
eventually enter the loss cone, although not all manage to do so in a Hubble time.
These considerations have led to a conclusion that the loss cone should remain full 
in axisymmetric and especially triaxial systems.

\section{$N$-body simulations and their limitations}

Simulations of binary MBH evolution initially focused on the spherical case, starting 
the evolution from a near-equilibrium stellar system with two embedded black holes 
before or just after the formation of the binary. 
At present, it seems impossible to evolve an $N$-body system with a number of particles 
comparable to that of a real galactic nucleus ($N_\star\gtrsim 10^8$), therefore all studies 
to date used a much smaller number of particles $N \lesssim 10^6$ and extrapolated 
the results to a real galaxy.
From the analytical estimates outlined above, it is clear that the evolution of the binary 
MBH in a real galaxy of spherical shape should occur in an empty-loss-cone regime. 
In this case, we expect the hardening rate of the binary (proportional to the repopulation 
rate of the loss cone) to scale with the number of particles in the simulation as 
$T_\mathrm{rel}^{-1} \propto N^{-1}$. 
However, it is worth noting that the empty-loss-cone regime is not so easy to achieve -- 
$N$ should be large enough so that even a very small fraction of the relaxation time is 
still much longer than the dynamical time \citep[see][Figure 15, for quantitative estimates 
of the minimum required $N$]{HarfstGMSPB2007}.
In the early studies \citep{QuinlanHernquist1997,MilosMerritt2001}, this was not possible 
and the hardening rate was almost independent of $N$ -- a signature of the full-loss-cone 
regime. Later simulations with a larger $N$ have confirmed that the hardening rate drops 
with $N$ if it is large enough \citep{MakinoFunato2004,BerczikMS2005,MerrittMS2007}.
At the same time, studies that considered a flattened model for the nucleus 
\citep{BerczikMSB2006} or a merger simulation that resulted in a non-spherical 
remnant \citep{PretoBBS2011,KhanJM2011} indicated no dependence of the hardening rate on $N$, 
which was found to be rather close to the full-loss-cone hardening rate. 
It was inferred that the departure from spherical symmetry was strong enough that 
the enhanced rate of angular momentum mixing sustains a full loss cone.

\begin{figure}
\includegraphics[scale=0.93]{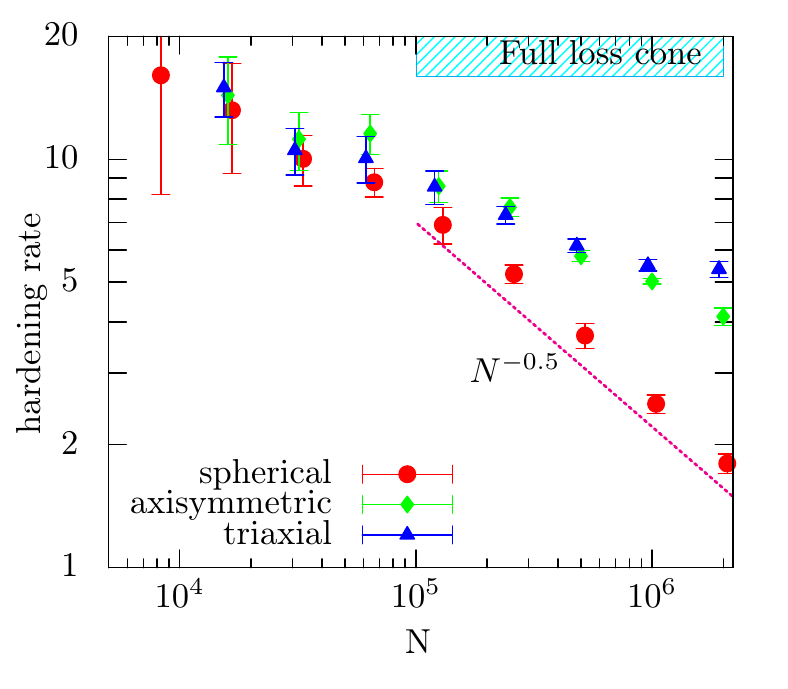}
\caption{Binary hardening rates $d(a^{-1})/dt$ as functions of the number of particles 
$N$ in simulations done in three geometries: spherical (red circles), axisymmetric 
(green squares) and triaxial (blue triangles). This figure is an updated version of the one 
presented in \citet{VasilievAM2014}, which in additions includes the results of simulations 
with $N=2\times10^6$.
It demonstrates that in all three cases the hardening rate does depend on $N$, 
although less so for non-spherical systems; there is a hint for levelling off at the highest 
$N$ in the triaxial case. In all three cases it remains substantially lower than 
the full-loss-cone rate for high enough $N$. }  \label{fig:hardening_rate}
\end{figure}

However, it remains to be proved that the apparent $N$-independence of the hardening rate 
in merger simulations is the result of non-spherical shape and not of some other physical 
process, such as an initially enhanced fraction of stars on radial orbits, or incomplete 
mixing of the large-scale structure of the merger remnant that could create time-dependent 
variations in the potential, again enhancing the angular momentum mixing.
Recently, two papers attemted a ``controlled experiment'' by exploring the binary evolution
in a system constructed to be initially in equilibrium. 
Rather disappointingly, they have reached conflicting conclusions: while \citet{KhanHBJ2013} 
have found the hardening rate in their axisymmetric models to be $N$-independent, 
\citet{VasilievAM2014} have shown the rate to depend on $N$ quite significantly in 
the axisymmetric and even triaxial geometry, although not as strongly as in the spherical 
case (Figure~\ref{fig:hardening_rate}). The reasons for this discrepancy remain obscure, 
but may have something to do with the different methods used for preparing the initial 
conditions for the models: ``adiabatic squeezing'' technique \citep{HolleySigurdsson2006} 
in the former paper and Schwarzschild modelling in the latter one.

%%%%%%%%%%%%%%
%%%%%%%%%%%%

The latter paper also have demonstrated that it is indeed very difficult to properly 
isolate the effect of loss-cone repopulation caused by non-spherical torques, because 
in an $N$-body simulation with a feasible number of particles ($N\lesssim 10^6$) 
it is strongly contaminated by the artificially enhanced two-body relaxation, thought 
to be unimportant in real galaxies. Thus a robust extrapolation of the result of 
simulation to much higher $N$ remains a very challenging task. For instance, can we 
trust the hints of levelling off of $N$-dependence of hardening rate in the triaxial 
model at $N\ge 10^6$ in Figure~\ref{fig:hardening_rate}? These simulations have also 
been conducted for a much shorter interval of time than is necessary to lead the binary 
into the GW-dominated regime, and it is not certain that the hardening rate stays 
constant with time. Therefore, unless we can afford much higher values of $N$ and longer
simulation intervals, the answer to the question whether the final-parsec problem 
exists in non-spherical galaxies would remain elusive. It is clear that at present 
time conventional $N$-body simulations cannot reach the required regime in which 
collisional (relaxation) effects are minor compared to collisionless (non-spherical) ones.

\section{The novel Monte Carlo method}

Fortunately, there exist a technique that allows to reliably model the collisional evolution 
of stellar systems with much cheaper computational demand than a direct $N$-body simulation. 
This is the Monte Carlo method of stellar dynamics, pioneered by \citet{SpitzerHart1971} 
and \citet{Henon1971}. It is based on the standard two-body relaxation theory, which 
predicts the statistical effect of many individual encounters on the trajectory of a given 
star, without the need to simulate these encounters explicitly. These methods have been 
continuously developed up to the present time and are successfully applied to various problems 
of dynamical evolution of star clusters \citep{GierszHHH2013,Pattabiraman2013} and galactic 
nuclei \citep{FreitagBenz2002}. The mentioned independent implementations are based on 
the H\'enon's formulation of the Monte Carlo method, which relies substantially on the 
assumption of spherical geometry; the alternative Spitzer's formulation is free from this 
assumption, although historically has been used only in a spherical case.

We have developed a novel Monte Carlo method based on Spitzer's work, which can be applied 
in any geometry and can explicitly distinguish the collisionless and collisional effects. 
The detailed description of the method is given in \citet{Vasiliev2014}, here we outline 
its main features.
The evolution of an $N$-body system is followed by a hybrid approach combining a temporally 
smoothed extension of the self-consistent field method \citep{HernquistOstriker1992} 
with a prescription for relaxation in terms of position-dependent velocity perturbations,
based on Spitzer's original formulation of the Monte Carlo method. 
The orbits of particles are followed in a smooth, non-spherical potential represented 
by a suitable basis-set expansion \citep[cf.][]{BrockampBK2014, MeironLHBS2014}, with 
a high-accuracy adaptive-timestep orbit integrator. After each integrator timestep, 
the particle velocity is perturbed using velocity diffusion coefficients computed from 
a spherical isotropic distribution function that approximates the true distribution of 
particles. The amplitude of this perturbation can be scaled to mimic the relaxation rate 
of a target stellar system that is composed of $N_\star$ stars; this number is not related 
to the actual number of particles in the simulation and can be set at will, in particular, 
turning off the relaxation entirely corresponds to $N_\star=\infty$.
Orbits of all particles are integrated in parallel, independently from each other, 
for an interval of time (``episode'') that can be comparable or even much longer than 
$T_\mathrm{orb}$, although should be $\ll T_\mathrm{rel}$. At the end of the episode, 
the system state (coefficients of expansion of the non-spherical potential, and 
the spherical approximation to the distribution function) is updated using trajectories 
of particles sampled during the entire episode.
The main advantages of the algorithm are the much lower computational demand than a full 
$N$-body simulation (it scales linearly with $N$, since the motion of each particle depends 
only on the mean-field potential and the global distribution function, but not on 
explicit interactions with other particles), the possibility of scaling the relaxation rate 
to the desired level independently from $N$, and the ability to deal with potentials of 
arbitrary shape.

The application of this method to the problem of binary MBH evolution is based on the 
conservation approach \citep{SesanaHM2007,MeironLaor2012}. During each episode, particles 
are moving in the combined potential of the stellar cluster and the time-dependent potential 
of the binary MBH. At each close encounter of the particle with the binary, we record 
the change of energy and angular momentum of the particle; at the end of the episode, these 
changes are summed up, and the orbit of the binary is adjusted using the conservation law 
for energy and angular momentum. This procedure captures most relevant dynamical processes 
in the nucleus, such as the depletion and repopulation of the loss cone, secondary slingshot, 
and change of shape of the stellar potential, although it does not account for the influence 
of Brownian motion of the binary.
In this preliminary study, we focus on equal-mass binaries on circular orbits and do not 
follow the evolution of the eccentricity, which is expected to remain low throughout 
the evolution for these conditions \citep[e.g.][]{Sesana2010}.

\vspace{-0.2cm}
\section{Results}

%%%%%%%%%%%%%%
\begin{figure}[t]
\includegraphics{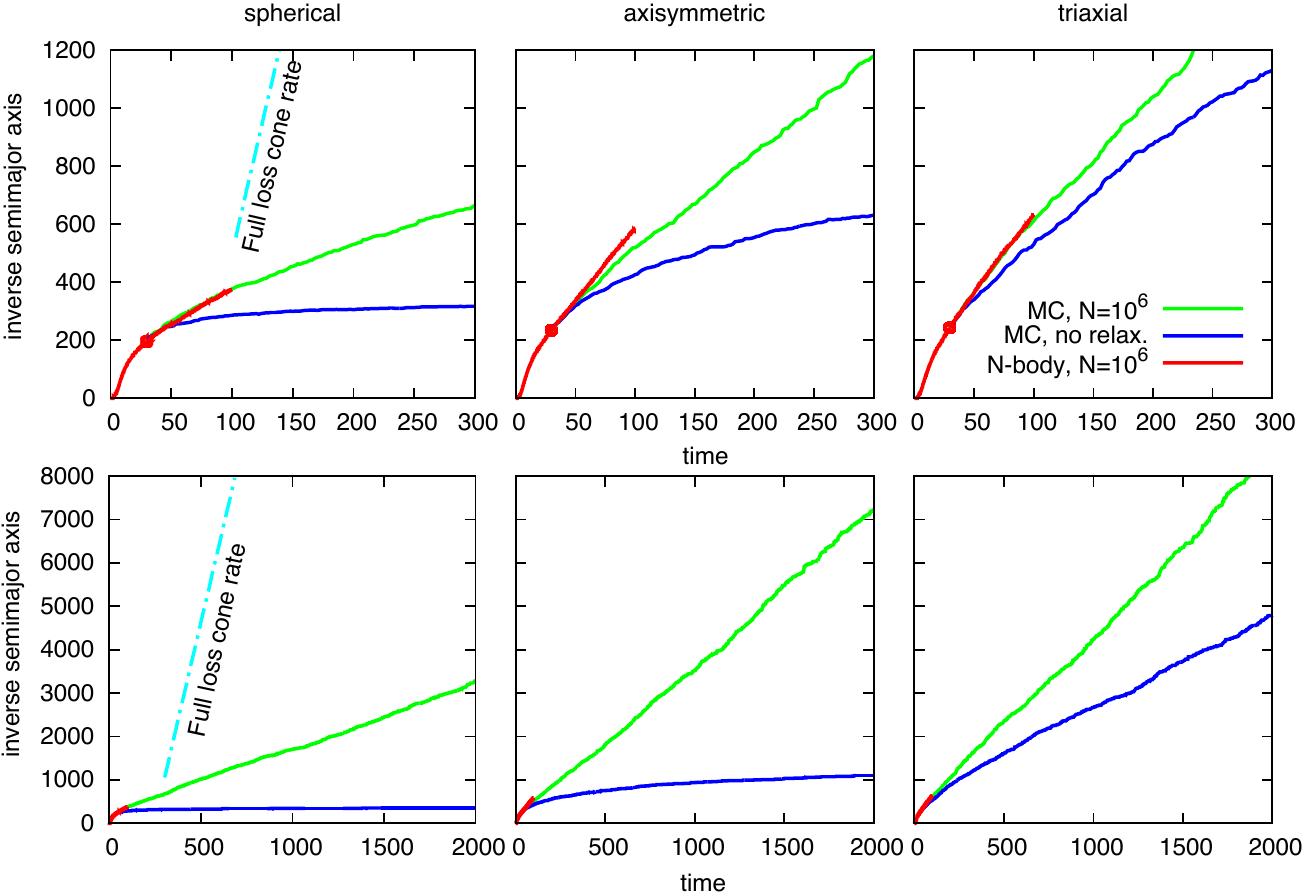}
\caption{The evolution of inverse binary semimajor axis $a^{-1}$ in $N$-body simulations 
of \citet{VasilievAM2014} (red curve, extending up to $t=100$) and in Monte Carlo 
calculations which included relaxation at the amplitude corresponding to the $N=10^6$ 
particle system from the $N$-body simulation (green curve, upper) or without relaxation 
(blue curve, lower). The thick red dot marks $t=30$ as the initial moment for the Monte 
Carlo calculations. Top row shows short-term evolution, which demonstrates the qualitative 
agreement between Monte Carlo and $N$-body simulations; bottom row is for long-term 
evolution. Left, central and right panels show three geometries -- spherical, axisymmetric, 
and triaxial. Only in the latter case the binary continues to shrink at an appreciable rate 
even without relaxation, although this rate is several times lower than the full-loss-cone 
rate shown as dot-dashed line. In the other two cases, the binary evolution without 
relaxation stalls at a separation just a few times smaller than the initial value of $a$, 
due to depletion of loss-cone orbits. On the other hand, with the relaxation rate 
corresponding to $N=10^6$ (state-of-the-art $N$-body simulations), there is considerable 
evolution in all three geometries, demonstrating that the $N$-body simulations cannot 
yet reach the dynamical regime relevant for real galaxies. }  \label{fig:hardness_evolution}
\end{figure}
%%%%%%%%%%%%

We start our Monte Carlo simulations from snapshots taken at $t=30$ in the $N$-body 
simulations of \citet{VasilievAM2014}, when the hard binary has already formed. 
We follow the evolution of the system in three cases (spherical, axisymmetric, and triaxial, 
with the last two systems having initial axis ratios of 1:1:0.8 and 1:0.9:0.8), with two 
choices of the relaxation rate: corresponding to a $N=10^6$ particle system (the same as 
the original simulations) and without relaxation (the situation relevant to real galaxies).
The results of this calculation, in terms of the evolution of inverse semimajor axis of 
the binary, are presented in Figure~\ref{fig:hardness_evolution}.

First of all, it confirms that the Monte Carlo method yields consistent results with 
direct $N$-body simulations, if the relaxation amplitude is scaled to the same level as 
in these simulations. Second, in the presence of relaxation the evolution continues with 
roughly the same hardening rate for much longer intervals of time, and there is indeed 
only a factor of two difference between three geometries at $N=10^6$. Third, and most 
importantly, it suggests that without relaxation, in the spherical and axisymmetric cases 
the evolution of binary semimajor axis $a$ slows down and eventually stalls at a distance 
only a few times smaller than the initial value of $a$, although it occurs later and at 
smaller distance in the axisymmetric case. By contrast, in the triaxial case the evolution 
continues without any relaxation at a rate which is just twice lower than the rate at 
$N=10^6$. 

The difference between the geometries is explained by the properties of orbits: 
as mentioned above, in the axisymmetric case there are more orbits that can enter the loss 
cone due to the variation in angular momentum caused by non-spherical torques than in 
the spherical case, where no such torques exist. However, the minimum value of angular 
momentum for any orbit is still non-zero in the axisymmetric geometry, and as the size of 
the loss cone shrinks proportionally with the binary separation, eventually all such orbits 
are depleted. By contrast, in the triaxial case the population of genuinely centrophilic 
orbits, all of them being able to achieve arbitrary low values of angular momentum, is 
large enough to sustain a steady influx of stars into the loss cone.
It is remarkable that the hardening rate in this case is about an order of magnitude lower 
than the full-loss-cone rate; nevertheless, scaled to real galaxies, it appears to be 
sufficient to drive the binary to the merger in a time much shorter than the Hubble time.

\section{Conclusions}

We have reviewed the evolution of binary massive black holes in gas-free galaxies, 
focusing on the final-parsec problem -- the difficulty to keep the shrinking rate of 
the binary at a high enough level for it to merge in a Hubble time. 
This problem is connected to the rate of repopulation of the loss cone -- region of 
phase space occupied by stars that are able to interact with the binary. In non-spherical 
galaxies this repopulation occurs both due to collisionless effects -- variations of 
angular momentum of stars due to torques from the global potential, and to collisional 
effects -- two-body relaxation. In real galaxies, collisional effects are supposed to 
play a very small role, however, in existing $N$-body simulations their contribution 
is greatly enhanced due to a limited number of particles. 

We have introduced a novel Monte Carlo method that is suitable for the dynamical evolution 
of non-spherical galactic nuclei and can explicitly adjust the relaxation rate. 
Preliminary calculations for equal-mass binary on a circular orbit suggest that 
in the absense of relaxation, the shrinking of the orbit stalls (the final-parsec problem 
exists) in spherical and axisymmetric systems, but not in the triaxial ones. 
In the future, we are going to apply the method for unequal-mass binaries, which are 
expected to develop substantial eccentricity that accelerates the transition to the 
GW-dominated regime \citep{Sesana2010,MeironLaor2012,Khan2012}. An open question is why 
merger simulations seem to imply the evolution rate comparable with the full-loss-cone 
rate, while our isolated systems never approach this rate for sufficiently high $N$; 
we are going to address it by running Monte Carlo simulations using merger remnants as 
initial conditions. Another interesting question is the rate of tidal disruptions of 
stars by either component of the binary at different evolutionary stages \citep[e.g.][]
{ChenMSL2009,WeggBode2011}; this process can also be included in the Monte Carlo code.

This work was partly supported by the National Aeronautics and Space Administration 
under grant no.\ NNX13AG92G. I thank David Merritt for comments on the manuscript, 
and Alberto Sesana for fruitful discussions.


\vspace*{-0.2cm}
\begin{thebibliography}{}
\setlength{\itemsep}{0.8pt}

\bibitem[Amaro-Seoane \etal(2010)]{AmaroSeoane2010}
Amaro-Seoane, P., Sesana, A., Hoffman, L., \etal\ 2010, \textit{MNRAS}, 402, 2308

\bibitem[Begelman \etal(1980)]{BegelmanBR1980} 
Begelman, M. C., Blandford, R. D., \& Rees, M. J. 1980, \textit{Nature}, 287, 307

\bibitem[Berczik \etal(2005)]{BerczikMS2005} 
Berczik, P., Merritt, D., \& Spurzem, R.  2005, \textit{ApJ}, 633, 680

\bibitem[Berczik \etal(2006)]{BerczikMSB2006} 
Berczik, P., Merritt, D., Spurzem, R., \& Bischof, H.  2006, \textit{ApJL}, 642, L21

\bibitem[Brockamp \etal(2014)]{BrockampBK2014}
Brockamp, M., K\"upper, A., Thies, I., Baumgardt, H., \& Kroupa, P.\ 2014, \textit{MNRAS}, 441, 150

\bibitem[Chatterjee \etal(2003)]{ChatterjeeHL2003} 
Chatterjee, P., Hernquist, L., \& Loeb, A.  2003, \textit{ApJ}, 592, 32

\bibitem[Chen \etal(2009)]{ChenMSL2009}
Chen, X., Madau, P., Sesana, A., \& Liu, F.-K.  2009, \textit{ApJL}, 697, L149

\bibitem[Dullo \& Graham(2014)]{DulloGraham2014}
Dullo, B., \& Graham, A.  2014, \textit{MNRAS}, 444, 2700

\bibitem[Ferrarese \& Ford(2005)]{FerrareseFord2005}
Ferrarese, L., \& Ford, E.  2005, \textit{Space Sci.\ Revs}, 116, 523

\bibitem[Frank \& Rees(1976)]{FrankRees1976}
Frank, J., \& Rees, M.  1976, \textit{MNRAS}, 176, 633

\bibitem[Freitag \& Benz(2002)]{FreitagBenz2002}
Freitag, M., \& Benz, W.  2002, \textit{A\&A}, 394, 345

\bibitem[Giersz et al.(2013)]{GierszHHH2013}
Giersz, M., Heggie, D., Hurley, J., \& Hypki, A.  2013, \textit{MNRAS}, 431, 2184

\bibitem[Harfst et al.(2007)]{HarfstGMSPB2007} 
Harfst, S., Gualandris, A., Merritt, D., Spurzem, R., Portegies Zwart, S., \& Berczik, P.\ 2007, \textit{New Astron.}, 12, 357

\bibitem[H\'enon(1971)]{Henon1971}
H\'enon, M., 1971, \textit{Ap\&SS}, 13, 284

\bibitem[Hernquist \& Ostriker(1992)]{HernquistOstriker1992} 
Hernquist, L., \& Ostriker, J.  1992, \textit{ApJ}, 386, 375

\bibitem[Holley-Bockelmann \& Sigurdsson(2006)]{HolleySigurdsson2006} 
Holley-Bockelmann, K., \& Sigurdsson, S.  2006, \textit{arXiv:astro-ph/0601520}

\bibitem[Iwasawa \etal(2006)]{IwasawaFM2006}
Iwasawa, M., Funato, Y., \& Makino, J.  2006, \textit{ApJ}, 651, 1059

\bibitem[Khan \etal(2011)]{KhanJM2011} 
Khan, F. M., Just, A., \& Merritt, D.  2011, \textit{ApJ}, 732, 89

\bibitem[Khan \etal(2012)]{Khan2012} 
Khan, F.~M., Preto, M., Berczik, P., Berentzen, I., Just, A., \& Spurzem, R.  2012, \textit{ApJ}, 749, 147 

\bibitem[Khan \etal(2013)]{KhanHBJ2013} 
Khan, F. M., Holley-Bockelmann, K., Berczik, P., \& Just, A.  2013, \textit{ApJ}, 773, 100

\bibitem[Kormendy \& Ho(2013)]{KormendyHo2013}
Kormendy, J., \& Ho, L.  2013, \textit{ARA\&A}, 51, 511

\bibitem[Lightman \& Shapiro(1977)]{LightmanShapiro1977}
Lightman, A., \& Shapiro, S.  1977, \textit{ApJ}, 211, 244

\bibitem[Lodato \etal(2009)]{LodatoNKP2009}
Lodato, G., Nayakshin, S., King, A., \& Pringle, J.  2009, \textit{MNRAS}, 298, 1392

\bibitem[Makino \& Funato(2004)]{MakinoFunato2004} 
Makino, J., \& Funato, Y.  2004, \textit{ApJ}, 602, 93

\bibitem[Meiron \& Laor(2012)]{MeironLaor2012}
Meiron, Y., \& Laor, A.  2012, \textit{MNRAS}, 422, 117

\bibitem[Meiron \etal(2014)]{MeironLHBS2014}
Meiron, Y., Li, B., Holley-Bockelmann, K., \& Spurzem, R.  2014, \textit{ApJ}, 792, 98

\bibitem[Merritt(2013)]{Merritt2013}
Merritt, D.,  2013, \textit{Classical \& Quantum Gravity}, 244005

\bibitem[Merritt \etal(2007)]{MerrittMS2007} 
Merritt, D., Mikkola, S., \& Szell, A.  2007, \textit{ApJ}, 671, 53

\bibitem[Merritt \& Poon(2004)]{MerrittPoon2004} 
Merritt, D., \& Poon, M.-Y.  2004, \textit{ApJ}, 606, 788

\bibitem[Milosavljevi{\'c} \& Merritt(2001)]{MilosMerritt2001} 
Milosavljevi{\'c}, M., \& Merritt, D.  2001, \textit{ApJ}, 563, 34

\bibitem[Milosavljevi{\'c} et al.(2002)]{MilosMRB2002} 
Milosavljevi{\'c}, M., Merritt, D., Rest, A., \& van den Bosch, F.  2002, \textit{MNRAS}, 331, L51

\bibitem[Milosavljevic \& Merritt(2003a)]{MilosMerritt2003a}
Milosavljevi{\' c}, M., \& Merritt, D. 2003a, in: J.~Centrella \& S.~Barnes (eds.), 
\textit{The Astrophysics of Gravitational Wave Sources}, AIP Conf.\ Proc.(Melville, NY: AIP), 686, p.\ 201

\bibitem[Milosavljevi{\'c} \& Merritt(2003b)]{MilosMerritt2003b} 
Milosavljevi{\'c}, M., \& Merritt, D.  2003b, \textit{ApJ}, 596, 860

\bibitem[Pattabiraman et al.(2013)]{Pattabiraman2013}
Pattabiraman, B., Umbreit, S., Liao, W.-K., Choudhary, A., Kalogera, V., 
Memik, G., \& Rasio, F.  2013, \textit{ApJS}, 204, 15

\bibitem[Perets \& Alexander(2008)]{PeretsAlexander2008}
Perets, H., \& Alexander, T.  2008, \textit{ApJ}, 677, 146

\bibitem[Poon \& Merritt(2001)]{PoonMerritt2001} 
Poon, M.~Y., \& Merritt, D.\ 2001, 549, 192

\bibitem[Preto \etal(2011)]{PretoBBS2011} 
Preto, M., Berentzen, I., Berczik, P., \& Spurzem, R.  2011, \textit{ApJL}, 732, L26

\bibitem[Ro\v skar \etal(2014)]{Roskar2014}
Ro\v skar, R., Mayer, L., Fiacconi, D., Kazantzidis, S., Quinn, T., Wadsley, J.
2014, \textit{arXiv:1406.4505}

\bibitem[Quinlan(1996)]{Quinlan1996} 
Quinlan, G.  1996, \textit{New Astron.}, 1, 35

\bibitem[Quinlan \& Hernquist(1997)]{QuinlanHernquist1997} 
Quinlan, G., \& Hernquist, L.  1997, \textit{New Astron.}, 2, 533

\bibitem[Sesana \etal(2007)]{SesanaHM2007} 
Sesana, A., Haardt, F., \& Madau, P.  2007, \textit{ApJ}, 660, 546

\bibitem[Sesana(2010)]{Sesana2010}
Sesana, A.  2010, \textit{ApJ}, 719, 851

\bibitem[Spitzer \& Hart(1971)]{SpitzerHart1971}
Spitzer, L., \& Hart, M.  1971, \textit{ApJ}, 164, 399

\bibitem[Vasiliev(2015)]{Vasiliev2014}
Vasiliev, E.  2015, \textit{MNRAS}, 446, 3150

\bibitem[Vasiliev \etal(2014)]{VasilievAM2014}
Vasiliev, E., Antonini, F., \& Merritt, D.  2014, \textit{ApJ}, 785, 163

\bibitem[Wegg \& Bode(2011)]{WeggBode2011}
Wegg, C., \& Bode, J. N.  2011, \textit{ApJL}, 738, L8

\bibitem[Yu(2002)]{Yu2002} 
Yu, Q.  2002, \textit{MNRAS}, 331, 935

\end{thebibliography}
\end{document}